# Synthesis and Structure of 1 D Na$_6$ Cluster Chain with Short Na-Na Distance: Organic like Aromaticity in Inorganic Metal Cluster

Snehadrinarayan Khatua, Debesh R. Roy, Pratim K. Chattaraj*, and Manish Bhattacharjee *

*Department of Chemistry, Indian Institute of Technology, Kharagpur 721 302, India*

*Fax: (+) 91-3222-282252 E-mail: mxb@iitkgp.ac.in*

**Abstract:** A unique 1D chain of sodium cluster containing (Na$_6$) rings stabilized by a molybdenum containing metalloligand has been synthesized and characterized and the DFT calculations show striking resemblance in their aromatic behaviour with the corresponding hydrocarbon analogues

Studies on polymeric metal compounds are receiving more and more attention in recent years due to their properties, which lie between isolated molecules and bulk material.[1] The major developments in this area are synthesis of conducting polymeric materials and metal cluster trapped within the channels of zeolite.[2,3] In recent years bimetallic compounds containing alkali metal clusters have been synthesized.[1a] For example, gallium phosphonates have been shown to form cages containing lithium,[1a] sodium, and potassium[4] ions with short metal-metal contacts. In the lithium compound the lithium ions are arranged in a finite unidimensional wire. These compounds are potential precursors for preparation of ion conductors and molecular sieves.[1a]

We have been interested in alkali metal complexes of molybdenum containing metalloligands and recently reported synthesis, structure, of [{Na$_4$(H$_2$O)$_4$(μ-H$_2$O)$_2$} ⊂ (Mo$_2$O$_5$L$_2$)$_2$], bearing rectangular Na$_4^{4+}$ cluster cation stabilized by the metalloligand, [LMoO$_2$(μ-O)MoO$_2$L]$^{2-}$ (LH$_2$ = {(3,5 – di – *tert* –butyl - 2 – hydroxybenzyl)amino}acetic acid) and the corresponding cesium compound, a helical 1D infinite chain of [Cs$_2$ (Mo$_2$O$_5$L$_2$). H$_2$O]$_n$.[5] We were interested in exploring methods for assembly of larger cluster. We report here synthesis and structure of a new metalloligand and its sodium complex [Na$_2$MoO$_3$L(H$_2$O)]$_n$ (**1**) {L = iminodiacetate}, which contains infinite one dimensional chain of hexagonal sodium ions with short Na – Na contacts and theoretical investigation (DFT calculation) of the stability and reactivity of that compound .

Our synthetic strategy is to use a tridentate, dibasic ligand (L), containing two carboxylate donor sites, that is expected to form the metalloligand, $[LMoO_3]^{2-}$. This polydentate metalloligand has a potential to coordinate alkali – metal cations through Mo=O oxygens and bridging carboxylate and self-assemble into various architectures (Scheme 1).

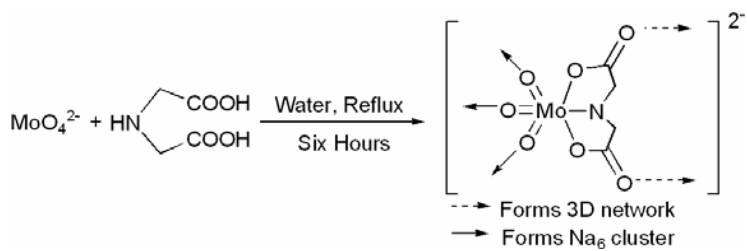

**Scheme 1**. Formation and coordination mode of the metaloligand

Accordingly, a reaction of $Na_2MoO_4 \cdot 2H_2O$ and iminodiacetic acid has been carried out in water and the compound **1** has been isolated in high yield[§]. The compound has been characterized by elemental analyses, thermogravimetric analysis (TGA) , and spectroscopic as well as a single crystal X-ray diffraction studies.

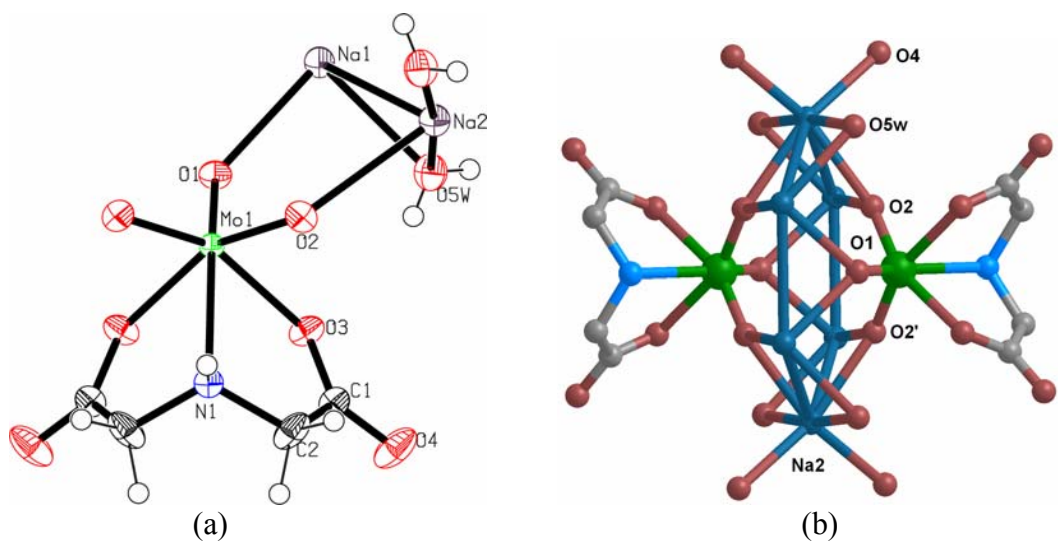

(a)            (b)

**Fig1** ORTEP view of **1**.Bond distances (Å) and angles(°): Mo(1)-O(1) 1.755(3), Mo(1)-O(2) 1.743(3), Mo(1)-O(3) 2.226(3), Na(1)-O(1) 2.352(3), Na(1)-O(2) 2.377(2), Na(2)-O(2) 2.485(3), Na(2)-O(4) 2.359(3), Na(2)-O(5W) 2.362(3), Na(1)-O(5W) 2.397(3) O(2)-Mo(1)-O(1) 107.02(10), O(2)-Mo(1)-O(3) 158.93(11), O(1)-Mo(1)-O(3) 86.11(11). (b) Part of the polymeric structure showing formation of $Na_6$ ring

X-ray diffraction analysis[¶] reveals that, the asymmetric unit of **1** contains half of the molybdenum coordinated iminodiacetate, two –oxo oxygens, two sodium, and one water as it sits on the inversion centre. The molybdenum atom is in a distorted octahedral environment and is coordinated to three –oxo oxygens and two carboxylate oxygens and a nitrogen (Figure 1a). Each of the two carboxylates of the ligand bridge one molybdenum atom and two sodium (Na2). Each of the three Mo=O oxygens bridges two sodium ions in a $\mu_3$ fashion (Figure 1b). In addition, the two sodium ions are bridged by one $\mu_2$-oxygen of water. Thus, each of the sodium ions is hexacoordinated and is in an distorted octahedral environment. The distinctive feature of the structure is the formation of hexagonal 1D chain of sodium ions (Figure 2a).

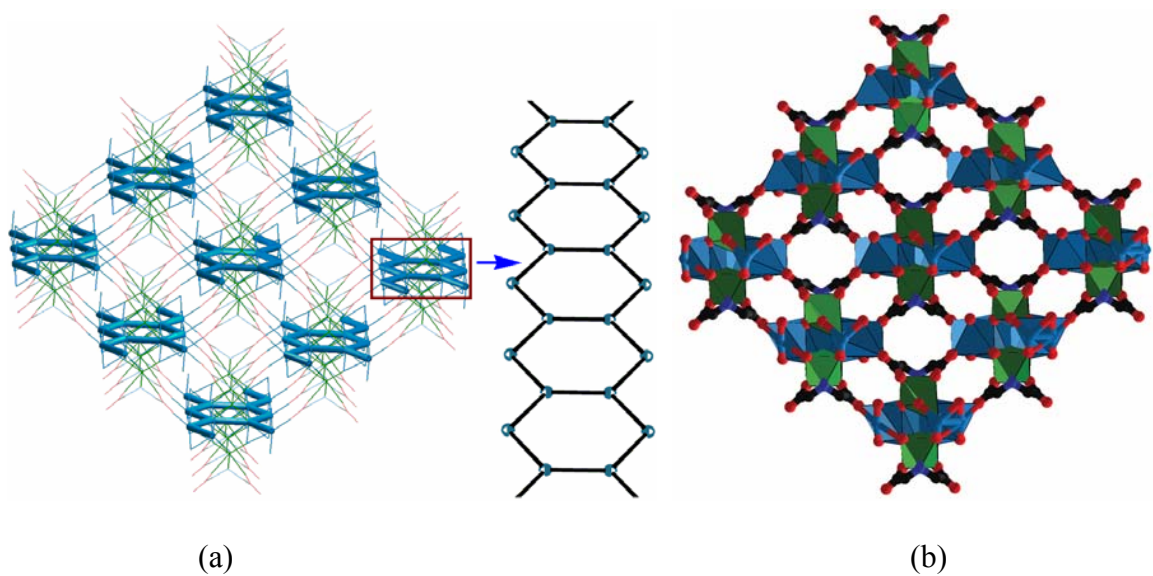

(a)                                                                 (b)

**Fig.2** (a) View of the polymeric network in **1** along c axis, where 1D chain of $Na_6$ units shown as thick blue line. (b) The 3D porous network in **1** along c axis depicted by OLEX[20]. Blue polyhedra represent ($NaO_6$) and the green octahedra represent ($MoO_5N$) units

The infinite polymeric chain is going towards the crystallograhic c axis. The observed Na….Na distances {Na1…Na1 = 3.074 (4) Å and Na1….Na2 = 3.2733 (18) Å} are distinctly shorter compared to that in elemental sodium (3.82 Å). The short Na⋯Na distances clearly show attractive Na – Na interaction in the cluster. In our earlier work on $Na_4^{4+}$ cluster[5] similar short Na⋯Na distance was observed. A few years back a tetrasodium dication, $Na_4^{2+}$, stabilized by two silyl(fluorosilyl)phosphonide with similar short Na – Na distances was reported.[6a] Recently,

Mehring and coworkers have reported shorter Na….Na distance in undecasodium decatrimethylsilanolate hydroxide.[6b] Roesky and coworkers, have also observed, similar but slightly longer Na….Na contacts in galophosphonates containing sodium ions.[4] Another, interesting feature of the 1D chain is that, the hexagonal ring ($Na_6$) is almost planer like benzene {∠Na1 Na1 Na2 = 123.83 (4) and ∠Na2 Na1 Na2 = 112.34 (8) }. It is interesting that the carboxylate groups in the metaloligand acts as a bridge between Mo and the Na centre of the neighbouring polymeric units and forms a porous 3D network (Figure 2b).

The observed geometry of the ($Na_6$) units in the 1D chain prompted us to investigate the characteristics of this cluster theoretically and look for a possible aromatic behavior. In recent years, all metal aromaticity has attracted a great deal of attention.[7] A few organometallic all metal

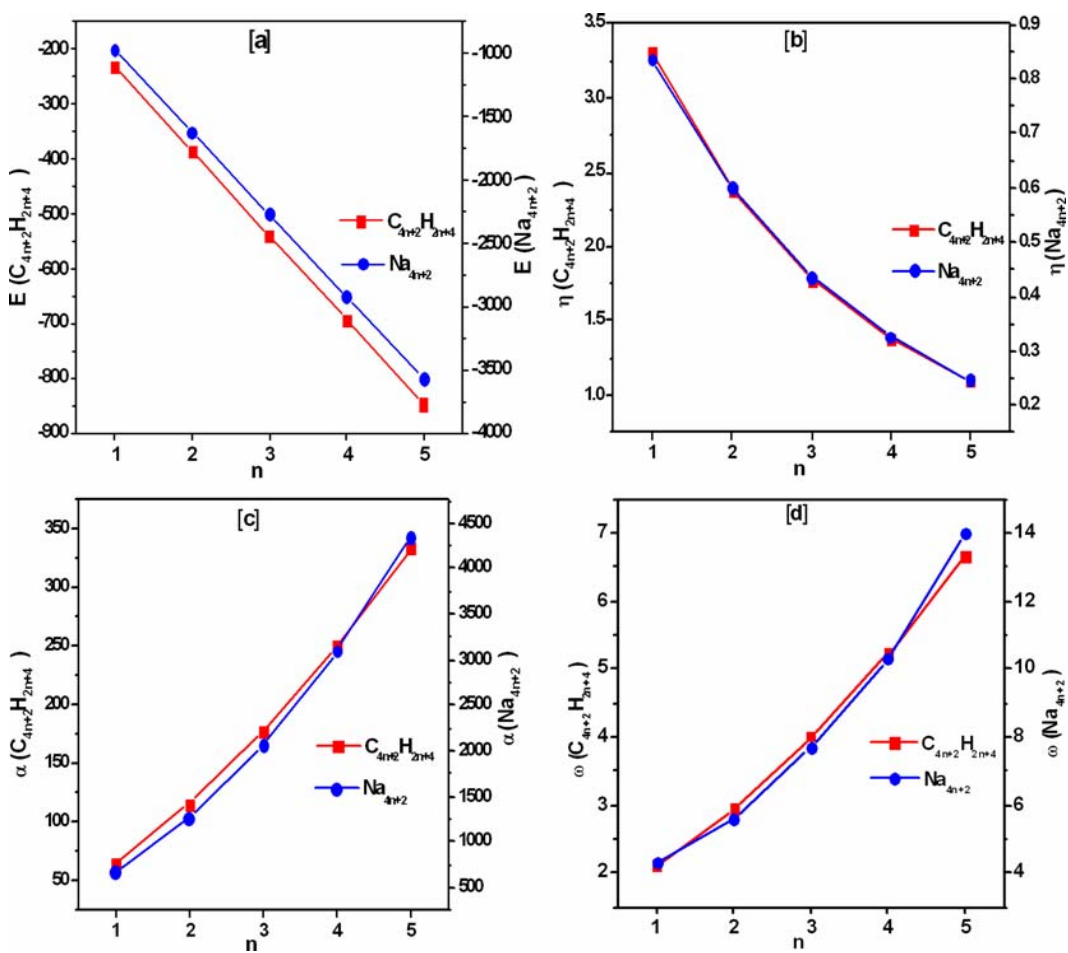

**Fig.3** (a) Energy (E, hartree), (b) Hardness (η, eV), (c) Polarizability (α, a.u.) and (d) Electrophilicity (ω, eV) profiles of the ($C_{4n+2}H_{2n+4}$) and ($Na_{4n+2}$), n = 1 - 5.

aromatic compounds have been reported.[7e] Apart form these report, the major focus has been on the gas phase synthesis and theoretical studies on aluminium cluster.[7] Starting from the experimental geometry of ($Na_{4n+2}$), n = 1 – 5, clusters single point calculations have been performed at the B3LYP/6-311+G* (for n = 1 – 4) and B3LYP/6-31G* (for n = 5) levels of theory. In the case of n = 5, the B3LYP/6-311+G* level of theory did not show any convergence. Thus in this case the B3LYP/6-31G* level of theory has been used. For all $C_{4n+2}H_{2n+4}$, n = 1-5, B3LYP/6-311+G* basis set is used. The electronic properties were also calculated at the B3LYP/LanL2DZ level of theory. It has been found that, except absolute values, the trends of

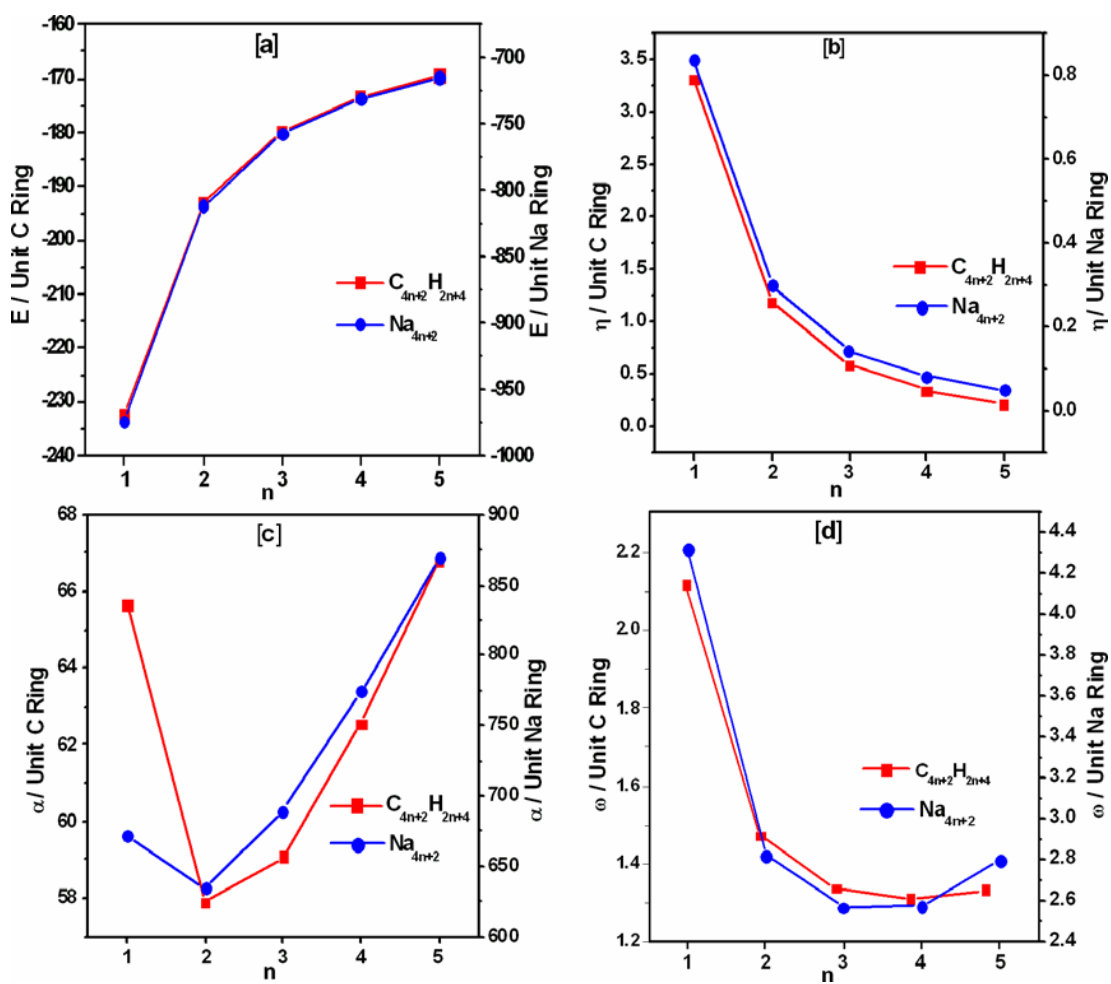

**Fig. 4** (a) Energy (E, hartree), (b) Hardness ($\eta$, eV), (c) Polarizability ($\alpha$, a.u.) and (d) Electrophilicity ($\omega$, eV) profiles per unit $C_6$ ring / $Na_6$ ring.

Various electronic properties remain same as those calculated using B3LYP/LanL2DZ level of theory. Various electronic properties like energy (E), hardness (η)[8], polarizability (α)[9], eletrophilicity (ω)[10], and nucleus independent chemical shift at the ring center {NICS(0)}[11] have been calculated using standard techniques.[12] Figure 3 presents the plots of E, η, α, and ω of ($Na_{4n+2}$), n = 1 to 5 clusters. The overall cluster gets softer, more polarizable and more electrophilic as n increases. Corresponding plots per ($Na_6$) unit are shown in Figure 4, which provides more transparent view of the stability and reactivity. As the chain length increases, its energy per ($Na_6$) unit increases and the hardness decreases as expected from the maximum hardness principle.[13] This implies that, the average reactivity increases with the increase in chain length. Except for one or two smallest clusters, the observed trend of variation in α and ω are as expected form the principles of the minimum polarizability[14] and electrophilicity.[15] Corresponding linear arenes (Figures 3 & 4) exhibit identical reactivity and stability patterns. These electronic structure principles have been shown to be adequate in explaining the aromatic and antiaromatic behaviour of all metal compounds[16] like $Al_4^{2-}$ and $Al_4^{4-}$ or the prototypical organic analogues like benzene and cyclobutadiene.[17]

In order to check the possible aromatic character of the $Na_{4n+2}$ clusters, their NICS(0) values have been calculated and compared with those values (B3LYP/6-311+G*) of the corresponding linear

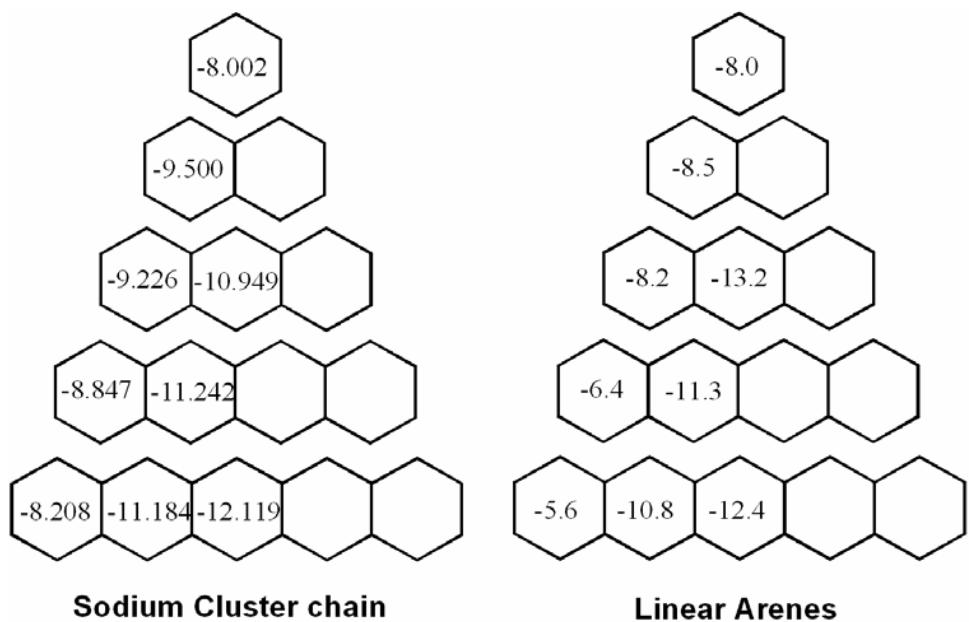

**Fig. 5** NICS (0) values (ppm) of each ($Na_6$) ring in the cluster chain [general formula ($Na_{4n+2}$)] and linear arenas at the 6-311 + G* level except for $Na_{22}$ which is calculated at the 6-31G* level

arenes.[18] The comparison is shown in Figure 5. It is interesting to note that, these ($Na_{4n+2}$), clusters exhibit striking resemblance, in their aromatic behaviour, with that of linear polyacenes. Not only their NICS (0) values are comparable, "the more reactive inner rings actually are more aromatic than the less reactive outer rings and even more aromatic than benzene itself"[19] for the linear arenes as well as ($Na_{4n+2}$) clusters.

In conclusion, a unique hexagonal chain of sodium has been synthesized and structurally characterized. The DFT calculations show that, the ($Na_6$) rings in the chain are highly aromatic in character and the NICS(0) values of the $Na_6$ rings are almost same as their polyacene analogues. The stability and reactivity pattern of the $Na_6$ rings also follow the same pattern as their organic analogues. Thus, the inorganic $Na_6$ chain shows organic like aromaticity.

**Acknowledgement**

We thank Department of Science & Technology, Government of India, New Delhi, for providing single crystal X-Ray facility under FIST grant and also for financial support.

**Notes and references**

§ An aqueous solution of $Na_2MoO_4.2H_2O$ (0.241 g, 1 mmol) was added to an aqueous solution of iminodiacetic acid (0.133 g, 1 mmol). The mixture was refluxed for about 6 hours, the resulting mixture was filtered and the filtrate was allowed to stand in air at room temperature. After six weeks colourless block-shaped crystals, suitable for X-ray diffraction, of **1** were obtained. Yield 80%. Elemental analysis calc (%) for $C_4H_9MoNNa_2O_9$(Mw =357.04): C 13.46, H 2.54, N 3.92, Mo 26.87, Na 12.88; found: C 13.44, H 2.49, N 3.89, Mo 26.39(gravimetric analyses), Na 13.16 (Flame Photometry). IR (KBr) ($v_{max}$ /cm$^{-1}$) 1660, 1400, 890, 840, 760. TGA: Total weight loss of 41.972 % from 170 ºC to 534 ºC is attributed to the loss of one coordinated $H_2O$, and one organic ligand (calculated 46.72 %). 4.748 % discrepancy is due to the one remaining oxygen which is used to form the residue. The residual weight 58.03 % (calculated: 57.685 %) corresponds to white coloured $Na_2MoO_4$.

¶ Crystal data for **1**: $C_4H_9MoNNa_2O_9$, Mw = 357.04, 298(2) K, Colourless Block (0.2 × 0.2 × 0.1 mm$^3$), monoclinic, C2/m, Z = 4, $a$ = 13.538(5) Å, $b$ = 14.980(3) Å, $c$ = 5.438(2) Å $\beta$ = 99.622(12)°, $V$ = 1087.4(6) Å$^3$, $\rho_{calcd}$ = 2.181 Mg/m$^3$, $2\theta_{max}$ = 49.88, $\lambda$($Mo_{k\alpha}$ = 0.71073 Å, $\omega$-

scan, 1039 measured, 997 unique ($R_{int}$ = 0.0266, $R_\sigma$ = 0.0065). R1 = 0.0303, wR2 = 0.0823 for 991 reflections with I > 2σ(I) and R1 = 0.0304, wR2 = 0.0824 for all data, max/min residual electron density = 1.145/-1.558. CCDC-608214. Data of **1** were collected on Bruker-Nonius Mach3 CAD4 X-ray diffractometer that uses graphite monochromated Mo Kα radiation (λ=0.71073 Å) by ω-scan method. No absorption correction was used. The structure was solved by Direct methods using the programme SHELXS-97 [21] and refined by least square methods on $F^2$ using SHELXL-97.[22] Non-hydrogen atoms were refined anisotropically and hydrogen atoms on C-atoms were fixed at calculated positions and refined using a riding model. Hydrogen atom connected to N atom were located in difference Fourier maps and refined isotropically with the N–H distances being restrained to 0.85(2) A°. Hydrogen atoms of the water molecule were also located in difference Fourier maps and refined isotropically.